\input harvmac
\lref\Z{G. Zwart, Nucl. Phys. B526 (1998) 378;\hfill\break
G. Aldazabal, A. Font, L.E. Ib\'a\~nez and G. Violero,
Nucl. Phys. B536 (1998) 29.}
\lref\BS{M. Bianchi and A. Sagnotti, Phys. Lett. {B247} (1990)
  517.}
\lref\zero{A. Sagnotti, {\it Some properties of open string theories},
hep-th/9509080; \hfill\break
A. Sagnotti, {\it Surprises in open string perturbation theory},
hep-th/9702093.} 
\lref\czero{C. Angelantonj, Phys. Lett. {B444} (1998) 309;\hfill\break 
C. Angelantonj, {\it Non-supersymmetric open string vacua}, hep-th/9907054.}
\lref\zerop{R. Blumenhagen, A. Font and D. L\"ust, {\it Tachyon-free
orientifolds  of type 0B strings in various dimensions}, hep-th/9904069.}
\lref\bkzero{R. Blumenhagen and A. Kumar, {\it A note on orientifolds
and dualities of type 0B string theory}, hep-th/9906234.}
\lref\ps{G. Pradisi and A. Sagnotti, Phys. Lett. {B216} (1989) 59.}
\lref\bsI{M. Bianchi, G. Pradisi
and A. Sagnotti, Nucl. Phys. {B376} (1992) 365;
\hfill\break
M. Bianchi, Nucl. Phys. {B528} (1998) 73;\hfill\break
E. Witten, J. High Energy Phys. {02} (1998) 006.}
\lref\carlo{Z. Kakushadze, G. Shiu and S.-H. Henry Tye, 
Phys.Rev. D58 (1998) 086001;\hfill\break
C. Angelantonj, {\it Comments on open-string orbifolds with a 
non-vanishing $B_{ab}$}, hep-th/9908064.}
\lref\DHSW{L.J. Dixon and J.A. Harvey, Nucl. Phys. {B274} (1986) 93;\hfill
\break N. Seiberg and E. Witten, Nucl. Phys. {B276} (1986) 272;
\hfill\break
L. Alvarez-Gaum\'e. P. Ginsparg, G. Moore and C. Vafa, Phys. Lett. {B171}
(1986) 155.}
\lref\cargese{A. Sagnotti, in Carg{\`e}se 87, Non-Perturbative Quantum Field
Theory, eds. G. Mack et al. (Pergamon Press, Oxford, 1988), p. 521.}
\lref\addsII{I. Antoniadis, G. D'Appollonio, E. Dudas and A. Sagnotti,
{\it Open Descendants of $Z_2\times Z_2$ Freely-Acting Orbifolds}, 
hep-th/9907184.}  
\lref\ztwo{M. Bianchi, PhD thesis, preprint ROM2F-92/13;\hfill\break
A. Sagnotti, {\it Anomaly Cancellations and Open-String
  Theories}, hep-th/9302099;\hfill\break
M. Berkooz and R.G. Leigh,  Nucl. Phys. {B483}
(1997).}
\lref\FS{W. Fischler and L. Susskind, Phys. Lett. {B171} (1986) 383;
Phys. Lett. {B173} (1986) 262.}
\lref\BG{O. Bergman and M. Gaberdiel, {\it Dualities of type 0
    strings}, hep-th/9906055.}
\lref\BD{J.D. Blum and K.R. Dienes, Phys. Lett. {B414} (1997) 260;
Nucl. Phys. {B516} (1998) 83.} 
\lref\ADS{I. Antoniadis, E. Dudas and A. Sagnotti, Nucl. Phys. {B544} 
(1999) 469.}
\lref\KT{ I.R. Klebanov and A.A. Tseytlin, Nucl. Phys. {B546}
    (1999) 155.}
\lref\FPS{D. Fioravanti, G. Pradisi and A. Sagnotti, Phys. Lett. B321
(1994) 349;\hfill\break
G. Pradisi, A. Sagnotti and Ya.S. Stanev, Phys. Lett. B345 (1995)
279; Phys. Lett. B 356 (1995) 230; Phys. Lett. B381 (1996) 97.}
\def\cA {{\cal A}}
\def\cM {{\cal M}}
\def\cK {{\cal K}}
\def\cT {{\cal T}}
\def\h {{1\over 2}}
\def\o {\over}
\def\th{\theta}
\def\La {\Lambda}
\def\la{\lambda}
\def\e{\epsilon}
\Title{\vbox{\rightline{\tt hep-th/9909010} \rightline{CPHT-S733.0999}}}
{\vbox{\centerline{On Non-tachyonic $Z_N\times Z_M$ Orientifolds} 
\vskip .2in 
\centerline{of Type 0B String Theory}}}
\vskip .2in 
\centerline{Kristin F\"orger}

\bigskip\centerline{\it Centre de Physique Th\'eorique}
\centerline{\it Ecole Polytechnique}
\centerline{\it 91128 Palaiseau Cedex, FRANCE}
\vskip 1in

\centerline{{\bf Abstract}}
We study open descendants of four dimensional 
$Z_N\times Z_M$ orbifolds of the non-supersymmetric 
type 0B string theory. An exhaustive analysis shows, that
using the crosscap constraint the only model 
for which one can project out the tachyon is the $Z_2\times Z_2$ orbifold.
For this case we explicitly construct the open string amplitudes.
The gauge group corresponding to the various inequivalent
Klein bottle projections turns out to be either symplectic or unitary.

\Date{9/99} 
\newsec{Introduction}
It is well known, that besides the five supersymmetric string theories
in ten dimensions there are also non-supersymmetric theories \DHSW, of
which the type 0B theory has recently attracted much attention.
In the AdS/CFT correspondence it has been considered as a
gravitational background in order to study
the dynamics of non-supersymmetric gauge theories in four dimensions
\KT. On the other hand, it was argued that type 0B theory corresponds
to a certain supersymmetry breaking orbifold of M-theory
\refs{\BD,\ADS,\BG} and further duality conjectures in lower
dimensions for non-supersymmetric theories have been discussed in \bkzero.

Type 0B theory can be obtained either by changing the GSO-projection
or by modding out type IIB theory by $(-1)^{F_s}$, where $F_s$ is the total
space-time fermion number. Its ten dimensional 
field content consists in a tachyon, a graviton, a dilaton and an 
antisymmetric tensor from the NS-NS sector as well as two scalars, 
two antisymmetric tensors and a four form from the R-R sector.
Since the fields in the R-R sector are doubled compared to type IIB
theory one expects to have twice the number of D-branes in type 0B theory.

Although the tachyon renders this theory unstable, 
it has been shown in \zero, that using the crosscap constraint \FPS \ 
non-tachyonic open descendants of type 0B theory can be constructed
along the lines of \refs{\BS,\zero,\cargese}. The resulting ten
dimensional theory contains a graviton, a dilaton, a scalar, an 
antisymmetric tensor and a self-dual four form in the unoriented 
closed string 
sector and a gauge vector in the adjoint of ${\rm U}(32)$ together
with Majorana-Weyl fermions in the $496\oplus\overline{496}$ 
representation in the open string sector. 
Recently this construction has been generalized to lower dimensional 
supersymmetry preserving $Z_N$ orientifolds in \refs{\czero,\zerop} 
and in \bkzero\ for a `non-supersymmetric' $Z_2$ orbifold.
It turns out that the only $Z_N$ orientifolds, for which
one can remove the tachyon is the $Z_2$ in $d=6$ and 
the $Z_3$ in $d=4$. Typically, in non-tachyonic type 0B descendants
one has to relax the dilaton tadpole, which can be cured by the
Fischler-Susskind mechanism \FS.

In this letter we study open descendants of four dimensional 
$Z_N\times Z_M$ orbifolds of type 0B theory.\foot{The construction 
of $Z_N\times Z_M$ orientifolds for type IIB theory has been done in \Z.}
An extensive analysis reveals that only for $N=M=2$ 
the tachyon can be projected out, thereby stabilizing the theory.
Therefore we focus on the explicit construction of open descendants
for the $Z_2\times Z_2$ orbifold.
The structure of open string 
amplitudes are similar to those of type IIB theory 
\refs{\ztwo,\addsII}, which contains $9$ branes as well as three different
sets of $5$ branes with gauge group ${\rm USp}(16)^{\otimes 4}$. 
In Section 2 we discuss compactifications on four dimensional
$Z_N\times Z_M$ orbifolds. We explicitly
construct the torus partition function for the $Z_2\times Z_2$
orbifold and derive the closed string spectrum. 
In Section 3 we study all possible inequivalent
Klein bottle projections, one of which projects out the tachyon.
We give the corresponding open string spectra for these cases.
\newsec{The parent type 0B theory}

We consider `supersymmetric' $Z_N\times Z_M$ orbifolds in $d=4$
of type 0B theory. The construction of the corresponding partition
functions reveals the fact, that generically the untwisted sector
contains a tachyon, which is directly related to the one in
ten dimensions. In addition, there also appear tachyons in the twisted 
sector except for the case $N=M=2$. For the same reasons as explained
in \refs{\czero,\zerop}, tachyons in the twisted sector cannot
be removed by any Klein bottle projection, since for a geometric
orbifold action these states do not appear in the Klein bottle amplitude.    
Therefore, in the following we restrict ourselves to the only possible
$Z_2\times Z_2$ orbifold, in which the crosscap constraint \FPS\ 
projects out the tachyon.

The orbifold generators act on the three internal tori 
$T^2\times T^2\times T^2$ as
$g=(1,-1,-1)$, $h=(-1,-1,1)$ and $f=g\cdot h=(-1,1,-1)$.
The starting point of the construction is the type 0B 
torus partition function
compactified down to four dimensions:
\eqn\torus{
\cT_{0B}=\Big[|O_8|^2+|V_8|^2+|S_8|^2+|C_8|^2\Big]\La_1\La_2\La_3\ ,}
where $\La_i$ with $i=1,2,3$ denote the lattice sums of the 
internal tori.
Decomposing the affine ${\rm SO}(8)$ characters $O_8$, $V_8$, $S_8$, $C_8$ 
into products of ${\rm SO}(2)$ characters $O_2$, $V_2$, $S_2$, $C_2$,
the $Z_2\times Z_2$ orbifold partition
function splits into four parts, which are associated to the four 
contributions in \torus.
Their structure is similar to the one of the supersymmetric case
studied in \refs{\ztwo,\addsII}. Thus we write 
\eqn\orbtor{
\cT=\cT_{\rm A}+\cT_{\rm B}+\cT_{\rm C}+\cT_{\rm D}\ }
with
$$
\eqalign{
\cT_{\rm A}&={1\o 4}\Big\{|A_{oo}|^2\La_1\La_2\La_2+
\Big(|A_{og}|^2\La_1+|A_{oh}|^2\La_3+|A_{of}|^2\La_2\Big)\Big|
{\th_3^2\th_4^2\o\eta^4}\Big|^2\cr
&+\Big[|A_{g o}|^2 \La_1+|A_{h o}|^2 \La_3+|A_{f o}|^2\La_2\Big]
\Big|{\th_2^2 \th_3^2\o \eta^4}\Big|^2\cr
&+\Big[|A_{g g}|^2\La_1+|A_{hh }|^2 \La_3+|A_{ff}|^2\La_2
\Big]\Big|{\theta_2^2\theta_4^2\o \eta^4}\Big|^2\cr
&+\Big[|A_{g h}|^2+|A_{gf}|^2+|A_{ hg}|^2+|A_{hf}|^2+
|A_{ fg}|^2+|A_{fh}|^2\Big]\Big|{\th_2^2\th_3^2\th_4^2\o \eta^6}\Big|^2 
\Big\}\ ,}
$$
where we introduced $16$ expressions $A_{ij}$ with 
$i,j=o,g,h,f$:
$$
\eqalign{
A_{io}&=\alpha_{io}+\alpha_{ig}+\alpha_{ih}+\alpha_{if}\ ,\quad \quad
A_{ig}=\alpha_{io}+\alpha_{ig}-\alpha_{ih}-\alpha_{if}\ ,\cr
A_{ih}&=\alpha_{io}-\alpha_{ig}+\alpha_{ih}-\alpha_{if}\ ,\quad\quad 
A_{if}=\alpha_{io}-\alpha_{ig}-\alpha_{ih}+\alpha_{if}\ ,}
$$
and $\alpha_{ij}$ are combinations of products of
 ${\rm SO}(2)$ characters, which
are $Z_2\times Z_2$ eigenstates with eigenvalue $\pm 1$.
We define similar quantities $B_{ij}$, $C_{ij}$ and $D_{ij}$,
depending on $\beta_{ij}$, $\gamma_{ij}$ and $\delta_{ij}$, respectively,
which are listed in the Appendix.
At the origin of the lattice we can rewrite \orbtor\ in terms of $64$
generalized characters (see Appendix):
$${
\cT^{(0)}=\sum_{k=1}^4\ \sum_{i=o,g,h,f} \Big(
|\chi_{\alpha,ik}|^2+|\chi_{\beta,ik}|^2+|\chi_{\gamma,ik}|^2+
|\chi_{\delta,ik}|^2\Big)\ ,}
$$
from which one can easily read off the spectrum.
Expansion of the characters of the untwisted sector yields: 
$${
\cT_{\rm untw}^{(0)}\sim {\rm tachyon}+|V_2|^2+3\  |2\ 
O_2|^2+8\ |S_2+C_2|^2+{\rm massive}
\ ,}
$$
and therefore the low lying states consist in one tachyon, 
one graviton, $30$  scalars and $8$ vectors.

The massless contributions of the $g$, $h$ and $f$ twisted sector
are
$${
\cT_{\rm tw}^{(0)}\sim 16\cdot 3\ \Big[4\  |O_2|^2+2\  |S_2|^2+
2\  |C_2|^2\Big]\,} 
$$
from which we can extract $192$ scalars and $96$ vectors. 
All together the closed string spectrum consists in 
one tachyon, one graviton, $222$ scalars and $104$ Abelian vectors.
\newsec{Open descendants}
In general the choice of signs in the Klein bottle is not unique
\zero. One can choose three inequivalent sets of signs
in the direct Klein bottle amplitude, which correspond to projections
$\Omega_1=\Omega$, $\Omega_2=\Omega (-1)^{F_{sR}}$ and 
$\Omega_3=\Omega (-1)^{F_R}$, 
where $F_{sR} (F_R)$ is the right-moving space-time 
(worldsheet) fermion number. 
The various cases can be expressed as:
$$
\eqalign{
\cK&={1\o 8}\Big\{ [\e_A A_{oo}+\e_B B_{oo}+\e_C C_{oo}+\e_D D_{oo}]
\Big(P_1P_2P_3+P_1W_2W_3+W_1P_2W_3+W_1W_2P_3\Big)\cr
&+2\cdot 16\Big([\e_A' A_{go}+\e_B' B_{go}+\e_C' C_{go}+\e_D' D_{go}] 
(P_1+W_1)(\phi_s+\phi_c)^2\cr
&+[\e_A' A_{ho}+\e_B' B_{ho}+\e_C' C_{ho}+\e_D' D_{ho}] 
(P_3+W_3)(\phi_s+\phi_c)^2\cr
&+[\e_A' A_{fo}+\e_B' B_{fo}+\e_C' C_{fo}+\e_D' D_{fo}] (P_2+W_2)(\phi_s+\phi_c)^2\Big)\Big\}\,}
$$
where $P_i(W_i)$ stands for the lattice sum of the KK momenta (winding
modes) and $\phi_s$ and $\phi_c$ are defined in the appendix. 
Moreover, we introduced signs $\e_A,\e_B,\e_C,\e_D=\pm 1$ in the
untwisted sector and $\e_A',\e_B',\e_C',\e_D'=\pm 1$ in the
twisted sector.
The natural choice of signs, which corresponds to the $\Omega_1$
projection, is to associate plus signs to bosons 
$\e_A,\e_A',\e_B,\e_B'=1$ and minus signs to fermions 
$\e_C,\e_C',\e_D,\e_D'=-1$. We will denote the corresponding 
Klein bottle by $\cK_1$.
The second choice $\Omega_2$, corresponds 
to all plus signs $\e_A,\e_B,\e_C,\e_D=1$ and $\e_A',\e_B',\e_C',\e_D'=1$.
The most interesting choice $\Omega_3$, 
has $\e_A,\e_C,\e_B',\e_D'=-1$ and $\e_B,\e_D,\e_A',\e_C'=1$, 
which will project out the closed string tachyon.
The different projections give rise to three different 
open string spectra.

The  unoriented closed spectrum is obtained from $\h
\cT+\cK$. In the first case the low-lying excitations contain the
tachyon, a graviton and $18$ scalars from the untwisted sector and
each of the $16$ fixed points in the $g,h$ and $f$-twisted sector
contributes with $4$ additional scalars. 
The second Klein bottle projection leaves the tachyon, a graviton, $8$
vectors and $18$ scalars in the untwisted sector and 
$64$ scalars and $32$ vectors for each $g$, $h$ and
$f$ twisted sector.
Finally, the third Klein bottle projects out the closed string tachyon
and the unoriented closed string spectrum
comprises one graviton, $4$ vectors and $18$ scalars from the 
untwisted sector
as well as $32$ scalars and $16$ vectors for each $g,h$ and 
$f$-twisted sector.

A modular $S$ transformation turns the direct-channel Klein bottle $\cK$
into the transverse channel amplitude $\tilde\cK$. 
At the origin of the lattice sum, we find that the
coefficients of the characters are perfect squares, which is required
for consistency \refs{\FPS,\BS}.
In the first case, we thus get 
$$
\eqalign{
\tilde\cK_1^{(0)}&={2^6\o 8}\Big\{
\chi_{\beta,o1}\Big(\sqrt{v_1v_2v_3}+\sqrt{{v_1\o v_2 v_3}}
+\sqrt{{v_3\o v_1 v_2}}+\sqrt{{v_2\o v_1 v_3}}\Big)^2\cr
&+\chi_{\beta,o2}\Big(\sqrt{v_1v_2v_3}+\sqrt{{v_1\o v_2 v_3}}
-\sqrt{{v_3\o v_1 v_2}}-\sqrt{{v_2\o v_1 v_3}}\Big)^2\cr
&+\chi_{\beta,o3}\Big(\sqrt{v_1v_2v_3}-\sqrt{{v_1\o v_2 v_3}}
+\sqrt{{v_3\o v_1 v_2}}-\sqrt{{v_2\o v_1 v_3}}\Big)^2\cr
&+\chi_{\beta,o4}\Big(\sqrt{v_1v_2v_3}-\sqrt{{v_1\o v_2 v_3}}
-\sqrt{{v_3\o v_1 v_2}}+\sqrt{{v_2\o v_1 v_3}}\Big)^2\Big\}\ ,}
$$
where $v_i$ are the volume factors of the three tori.
The amplitudes for $\tilde\cK_2$ and $\tilde\cK_3$ 
are readily obtained from $\tilde\cK_1$ by
replacing the characters $\chi_{\beta,ok}$ by  $\chi_{\alpha,ok}$
and $-\chi_{\gamma,ok}$, respectively.

The open unoriented sector is constructed from the annulus and 
M\"obius amplitudes $\cA+\cM$. For simplicity,
we only write the massless contributions to the amplitudes.
The transverse annulus for all three cases 
can be summarized as:
$$
\tilde\cA^{(0)}=\e_A\tilde\cA_A^{(0)} +\e_B\tilde\cA_B^{(0)} 
+\e_C\tilde\cA_C^{(0)}+\e_D\tilde\cA_D^{(0)}\ ,
$$
with the signs $\e_A,\e_B,\e_C$ and $\e_D$ as defined above, according to the three
models. Each term can be arranged into four characters which are
multiplied by independent squared reflection coefficients, e.g. for
$\tilde\cA_A^{(0)}$ we get: 
$$
\eqalign{  
\tilde\cA_A^{(0)}&={2^{-6}\o 8}\Big\{
\chi_{\alpha,o1} \Big(N_A\sqrt{v_1 v_2 v_3}
+D_{A1}\sqrt{{v_1\o v_2  v_3}}+ D_{A3} \sqrt{{v_3\o v_1 v_2}}
+D_{A2}\sqrt{v_2\o v_1 v_3}\Big)^2\cr 
&+\chi_{\alpha,o2} \Big(N_A\sqrt{v_1 v_2 v_3}
+D_{A1}\sqrt{v_1\o v_2 v_3}- D_{A3} \sqrt{{v_3\o v_1 v_2}}
-D_{A2}\sqrt{v_2\o v_1 v_3}\Big)^2\cr 
&+\chi_{\alpha,o3} \Big(N_A\sqrt{v_1 v_2 v_3}
-D_{A1}\sqrt{v_1\o v_2 v_3}+D_{A3} \sqrt{{v_3\o v_1 v_2}}
-D_{A2}\sqrt{v_2\o v_1 v_3}\Big)^2\cr 
&+\chi_{\alpha,o4} \Big(N_A\sqrt{v_1 v_2 v_3}
-D_{A1}\sqrt{v_1\o v_2 v_3}- D_{A3} \sqrt{{v_3\o v_1 v_2}}
+D_{A2}\sqrt{v_2\o v_1 v_3}\Big)^2\Big\} \ ,}
$$
and for $\tilde\cA_B^{(0)}$, $\tilde\cA_C^{(0)}$ and $\tilde\cA_D^{(0)}$
the characters have to be replaced by $\chi_{\beta,ok},
\chi_{\gamma,ok}$ and $\chi_{\delta,ok}$ and the sum of Chan-Paton 
charges $\{N_A,D_{Ar}\}$ $(r=1,2,3)$ by $\{N_B,D_{Br}\}$,
$\{N_C,D_{Cr}\}$ and $\{N_D,D_{Dr}\}$, respectively.
Notice, that in the present case only characters of the untwisted sector are involved in
the transverse annulus. This is due to the fact \refs{\FPS,\BS}, 
that only those characters are allowed to appear in $\tilde\cA$, 
which fuse into the identity together with their anti-holomorphic 
partners in the closed string GSO.

For the first model 
the transverse M\"obius amplitude at the origin of
the lattice reads
$$
\eqalign{
\tilde\cM_1^{(0)}&=-{1\o 4}\Big\{
\hat{B}_{oo}\Big[ N_B \ v_1 v_2 v_3+D_{B1}\ {v_1\o v_2 v_3}+
D_{B3}\  {v_3\o v_2 v_1}+D_{B2} \ {v_2\o v_1 v_3}\Big](\hat{\phi}_o+
\hat{\phi}_v)^3\cr
&+\hat{B}_{og}\Big[ (N_B+D_{B,1})\ v_1+(D_{B2}+D_{B3})\ {1\o v_1}\Big]
(\hat{\phi}_o+\hat{\phi}_v)(\hat{\phi}_o-\hat{\phi}_v)^2\cr
&+\hat{B}_{oh}\Big[ (N_B+D_{B3})\ v_3+(D_{B1}+D_{B2})\ {1\o v_3}\Big]
(\hat{\phi}_o+\hat{\phi}_v)(\hat{\phi}_o-\hat{\phi}_v)^2\cr
&+\hat{B}_{of}\Big[ (N_B+D_{B2})\ v_2+(D_{B1}+D_{B3})\ {1\o v_2}\Big]
(\hat{\phi}_o+\hat{\phi}_v)(\hat{\phi}_o-\hat{\phi}_v)^2\Big\}\ ,}
$$
and for $\tilde\cM_2^{(0)}$ and $\tilde\cM_3^{(0)}$
one has to replace $\hat{B}_{ij}$ by $\hat{A}_{ij}$ and
$\hat{C}_{ij}$ and the sum of Chan-Paton charges
$\{N_B,D_{Br}\}$ by $\{N_A,D_{Ar}\}$ and $\{N_C,D_{Cr}\}$,
respectively. All quantities have to be written in terms of
real `hatted' characters \refs{\zero,\BS}  and 
$\phi_o$ and $\phi_v$ are defined in the Appendix.

A modular $P=T^{\h}S\ T^2 S\  T^{\h}$ transformation turns the transverse 
channel M\"obius amplitude
into the direct channel amplitude, resulting in
\eqn\dirmoeb{
\cM_1^{(0)}={1\o 4}
\Big(\hat{\chi}_{\beta,o1}-\hat{\chi}_{\beta,o2}-\hat{\chi}_{\beta,o3}-
\hat{\chi}_{\beta,o4}\Big)\Big[N_B+D_{B1}+D_{B2}+D_{B3}]\ ,}
and similarly for $\cM_{2}^{(0)}$ with 
$(-\hat{\chi}_{\alpha,oi})$ and $\{N_A,D_{Ar}\}$
and  $\cM_{3}^{(0)}$ with $\hat{\chi}_{\gamma,oi}$ and $\{N_C,D_{Cr}\}$.

The tadpole conditions can be extracted from $\tilde\cK^{(0)}$,
$\tilde\cA^{(0)}$ and $\tilde\cM^{(0)}$ by setting to zero the
reflection coefficients of the massless characters.
One finds that
\eqn\tadI{\eqalign{
{\rm Model }\ 1 & :N_B=D_{B1}=D_{B2}=D_{B3}=64\ ,\cr 
{\rm Model }\ 3 & :N_C=D_{C1}=D_{C2}=D_{C3}=64\ .}}
For the second model there is no corresponding expression, 
since the characters which appear in all three transverse amplitudes
are either tachyonic or massive. In addition to the tadpole 
conditions \tadI\ one has to solve homogeneous tadpole conditions, 
which arise from additional massless characters in the transverse
annulus. They will be given later on during the discussion of each model.

The above tadpole condition fix the rank of the gauge group and therefore
the gauge group of the second model has no definite rank. 
Switching on a non-trivial quantized NS-NS antisymmetric tensor
of rank $r$ for the internal lattice
reduces the rank of the gauge group by a factor of
$2^{r/2}$, as it was shown in \bsI\  for toroidal compactification and
in \carlo\  for orbifold compactification.  
In the present case, three different choices are allowed:
$r=2,4,6$, which will modify the tadpole condition   
resulting in $32,16,8$ on the right hand side of \tadI.

In order for the direct channel M\"obius amplitude to correctly 
(anti-)symmetrize the direct channel annulus, 
the coefficients $N_A,N_B,N_C,N_D$ and 
$D_{Ar},D_{Br},D_{Cr},D_{Dr}$ have to be
parameterized in terms of Chan-Paton charges:
\eqn\ndcoef{\eqalign{
N_A&=N_o+N_v-N_s-N_c \quad\ ,\quad
D_{Ar}=D_{or}+D_{vr}-D_{sr}-D_{cr}\cr
N_B&=N_o+N_v+N_s+N_c \quad\ ,\quad
D_{Br}=D_{or}+D_{vr}+D_{sr}+D_{cr}\cr
N_C&=N_o-N_v-N_s+N_c \quad\ ,\quad
D_{Cr}=D_{or}-D_{vr}-D_{sr}+D_{cr}\cr
N_D&=N_o-N_v+N_s-N_c \quad\ ,\quad
D_{Dr}=D_{or}-D_{vr}+D_{sr}-D_{cr}\ .}}
Inspection of the direct channel M\"obius amplitude \dirmoeb\ 
requires the gauge group to be symplectic for the first model 
and unitary for the second and third model.
In the following, let us discuss each model separately.

{\it Model 1:}
In this case the remaining tadpole conditions result in 
$$
N_o=N_v\ , \quad N_s=N_c\ ,
\quad D_{or}=D_{vr}\ , \quad
D_{sr}=D_{cr}
$$ 
for $r=1,2,3$.
As in the supersymmetric case, one has to rescale the Chan-Paton charges 
such that $N_i=2 \ n_i$ and $D_{ir}=2\  d_{ir}$  $(i=o,v,s,c)$,
in order to have a consistent particle interpretation of 
$\cA^{(0)}+\cM^{(0)}$.
The direct channel annulus amplitude then reads
\eqn\dann{\eqalign{
\cA^{(0)}_1&={1\o 2}
\sum_{k=1}^4\chi_{\beta,ok} \ \sum_{i=o,v,s,c} (n_i^2+\sum_{r=1}^3
d_{ir}^2)\cr
&+\sum_{k=1}^4\chi_{\alpha,o k}  \Big[n_o n_v+n_s n_c+
\sum_{r=1}^3 (d_{or} d_{vr}+d_{sr} d_{cr})\Big]\cr
&- \sum_{k=1}^4\chi_{\gamma,o k}  \Big[n_o n_c+n_v n_s+
\sum_{r=1}^3 (d_{or} d_{cr}+d_{vr} d_{sr})\Big]\cr
&- \sum_{k=1}^4\chi_{\delta,o k}  \Big[n_o n_s+n_v n_c+
\sum_{r=1}^3 (d_{or} d_{sr}+d_{vr} d_{cr})\Big]\cr
&+  \sum_{k=1}^4\chi_{\alpha,g k}  \ \sum_{i\in\{o,v,s,c\}}(n_i d_{i1}+ 
d_{i2} d_{i3})
+ \sum_{k=1}^4\chi_{\beta,g k} 
\ \sum_{i,j\in\{o,v\},\{s,c\}\atop i\neq j}
(n_i d_{j1}+ d_{i2} d_{j3})\cr
&- \sum_{k=1}^4\chi_{\gamma,g k}  \ \sum_{i,j\in\{o,s\},\{v,c\}\atop i\neq j}
(n_i d_{j1}+ d_{i2} d_{j3})
- \sum_{k=1}^4\chi_{\delta,g k} 
 \ \sum_{i,j\in\{o,c\},\{v,s\}\atop
  i\neq j}(n_i d_{j1}+ d_{i2} d_{j3})\cr
&+ \sum_{k=1}^4\chi_{\alpha,h k}  \ \sum_{i\in\{o,v,s,c\}}(n_i d_{i3}+ 
d_{i1} d_{i2})
+ \sum_{k=1}^4\chi_{\beta,h k} 
\ \sum_{i,j\in\{o,v\},\{s,c\}\atop i\neq j}
(n_i d_{j3}+ d_{i1} d_{j2})\cr
&- \sum_{k=1}^4\chi_{\gamma,h k} 
 \ \sum_{i,j\in\{o,s\},\{v,c\}\atop i\neq j}
(n_i d_{j3}+ d_{i1} d_{j2})
- \sum_{k=1}^4\chi_{\delta,h k}  \ \sum_{i,j\in\{o,c\},\{v,s\}\atop
  i\neq j}
(n_i d_{j3}+ d_{i1} d_{j2})\cr
&+ \sum_{k=1}^4\chi_{\alpha,f k}  \ \sum_{i\in\{o,v,s,c\}}
(n_i d_{i2}+ d_{i1} d_{i3})
+ \sum_{k=1}^4\chi_{\beta,f k} 
\ \sum_{i,j\in\{o,v\},\{s,c\}\atop i\neq j}
(n_i d_{j2}+ d_{i1} d_{j3})\cr
&- \sum_{k=1}^4\chi_{\gamma,f k}  \ \sum_{i,j\in\{o,s\},\{v,c\}\atop i\neq j}
(n_i d_{j2}+ d_{i1} d_{j3})
- \sum_{k=1}^4\chi_{\delta,f k}  \ \sum_{i,j\in\{o,c\},\{v,s\}\atop
  i\neq j}
(n_i d_{j2}+ d_{i1} d_{j3})\ ,}}
where e.g. the notation $\sum _{i\neq j}d_{i1} d_{j2}$ for $i,j\in \{o,v\}$ 
stands for $d_{o1} d_{v2}+d_{v1} d_{o2}$.
The above annulus amplitude together with \dirmoeb\ leads to 
a symplectic gauge group:
$${
\Big[{\rm USp}(16-n_s)\otimes {\rm USp}(n_s)\Big]^{\otimes 2}\Big|_{9}
\otimes\prod_{r=1}^3 
\Big[{\rm USp}(16-d_s)\otimes {\rm USp}(d_s)
\Big]^{\otimes 2}\Big|_{5_r}\ .}
$$
If we choose $n_s=d_s=0$ the resulting theory turns out to be
purely bosonic with gauge group ${\rm USp}(16)^{\otimes 8}$
for  $D9, D9'$ branes as well as for $D5_r, D5_r'$ 
$(r=1,2,3)$ branes and the spectrum comprises
a tachyon in bi-fundamental representations and
massless scalars in the antisymmetric as well as combinations of
bi-fundamental representations of the gauge group.

The choice $n_s=d_s=8$ is particularly interesting, since
we get a vector in the adjoint of 
${\rm USp}(8)^{\otimes 16}$ and the following matter content:
$$
\eqalign{
{\rm tachyon}\ & :\bigoplus_{i=1}^8 (8_{2i-1},8_{2i})\cr
6 \phi\ &:\bigoplus_{i=1}^{16} \ 28_i\cr
4(\la_{\alpha}+\la_{\dot\alpha})\ &:\bigoplus_{i=1}^3
(8_{4i+1} \oplus 8_{4i+2}\ ,\ 8_{4i+3}\oplus 8_{4i+4})}
$$
in the untwisted sector, and
$$
\eqalign{
2\phi \ & : \bigoplus_{k=1}^3\bigoplus_{i=1}^{8-2k} 
(8_{2i-1}\oplus 8_{2i}\ ,\ 8_{2i+4k-1}\oplus 8_{2i+4k})\cr
(\la_{\alpha}+\la_{\dot\alpha})\ &:\bigoplus_{k=1}^3\bigoplus_{i=1}^{7-2k} 
(8_{2i-1}\oplus 8_{2i}\ ,\ 8_{2i+4k+1}\oplus 8_{2i+4k+2})\cr
&\oplus \bigoplus_{i=1}^3
(8_{4i-1}\oplus 8_{4i}\ ,\ 8_{4i+1}\oplus 8_{4i+2})}
$$
in the twisted sector.
Here the index of the fundamental or
antisymmetric representations
stands for the $i$-th factor of the gauge group.

{\it Model 2:}
The direct annulus for the second model
is obtained from \dann\  after the
replacement: 
$\chi_{\alpha,ik}\leftrightarrow \chi_{\beta,ik}$ and
$\chi_{\gamma,ik}\leftrightarrow \chi_{\delta,ik}$. 
Since the direct channel M\"obius amplitude only contains
characters $\chi_{\alpha,ok}$, the gauge fields transform under
a unitary gauge group. Therefore
one has to parametrize the charges as:
\eqn\chargesII{\eqalign{
N_o&=2 \ n\quad ,\  N_v= 2\  \bar n\quad , \ N_s= 2 \ m  \quad,\  N_c= 
2 \ \bar m  \cr 
D_{or}&=  2 \ d_r\quad ,\ D_{vr}=2\   \bar{d}_r \quad ,\  
D_{sr}= 2\  e_r\quad,\ 
D_{cr}=2\   {\bar  e}_r}}
Since the size of the gauge group is not fixed by the tadpole
conditions, one could set all charges to zero and would not have to
introduce any branes. 
However, for generic charges, one finds that a vector together
with six scalars transform in the adjoint of 
$$
{\rm U}(n)_{9}\otimes {\rm U}(m)_{9'}
\otimes\prod_{i=1}^3\Big[{\rm U}(d_i)_{5_i}
\otimes {\rm U}(e_i)_{5_i'}\Big]\ ,
$$
and the low lying states comprise 
$$
\eqalign{
{\rm tachyon}\ &:\bigoplus_{i=1}^8 (A_i\oplus {\bar A}_i)\cr
4(\la_{\alpha}+\la_{\dot\alpha})\ &:
\bigoplus_{i=1}^4\Big[(F_{2i-1},{\bar F}_{2i})\oplus (F_{2i-1}, F_{2i})
\oplus {\rm c.c.}\Big]\ ,}
$$
in the untwisted sector, and
$$
\eqalign{
2\phi\ &:  \bigoplus_{k=1}^3
\bigoplus_{i=1}^{8-2k}[(F_i,F_{2k+i})\oplus (F_i,{\bar
  F}_{2k+i})\oplus{\rm c.c.}]\cr
(\la_{\alpha}+\la_{\dot\alpha})\ &:
\bigoplus_{k=1}^3\bigoplus_{i=1}^{7-2k}
[(F_i,F_{2k+1+i})\oplus (F_i,{\bar F}_{2k+1+i})\oplus 
{\rm  c.c.}] \cr
&\oplus\bigoplus_{i=1}^3[(F_{2i},F_{2i+1})\oplus (F_{2i},{\bar
  F}_{2i+1})\oplus {\rm c.c.}]}
$$
in the twisted sector. Here, $A_i(F_i)$ denotes 
the antisymmetric (fundamental)
representation of the $i$-th factor of the gauge group.

{\it Model 3:}
This model is the most interesting one, because it is non-tachyonic.
Similarly to the previous case, the form of the M\"obius amplitude requires
a unitary gauge group. The direct channel annulus amplitude 
can again be obtained from \dann\ by
exchanging $\chi_{\alpha,ik}\leftrightarrow -\chi_{\delta,ik}$ and
$\chi_{\beta,ik}\leftrightarrow -\chi_{\gamma,ik}$ and
parameterizing the charges in \ndcoef \ similarly to model 2: 
$$
\eqalign{
N_o&=2 \ n\quad ,\  N_v= 2\  m\quad , \ N_c= 2\  \bar n  \quad,\  N_s= 
2\  \bar m  \cr 
D_{or}&=  2 \ d_r\quad ,\ D_{vr}=2\   e_r \quad ,\  D_{cr}= 2\ 
 {\bar d}_r\quad,\ 
D_{sr}=2 \  {\bar  e}_r}
$$
As usual in type 0B theory, one has to relax the dilaton tadpole condition,
which can be arranged  by the Fischler-Susskind mechanism \FS.
We can finally eliminate the open
string tachyon by setting $m=\bar m=e_r={\bar e}_r=0$.
The direct channel annulus amplitude then reads
$$
\eqalign{
\cA_{3}^{(0)}&=
 \sum_{k=1}^4 \chi_{\beta,ok}\Big[n\bar n+\sum_{r=1}^3 d_r{\bar
  d}_r\Big]
-\h \sum_{k=1}^4\chi_{\gamma,ok}\Big[n^2+{\bar n}^2
+\sum_{r=1}^3 (d_r^2+{\bar d}_r^2)\Big]\cr
&- \sum_{k=1}^4\chi_{\delta,gk}\Big[n d_1 +\bar n {\bar d}_1+ 
d_2 d_3 +{\bar d}_2 {\bar d}_3\Big]
+ \sum_{k=1}^4\chi_{\alpha,gk}\Big[n {\bar d}_1 +\bar n {d}_1+ 
d_2 {\bar d}_3 +{\bar d}_2 {d}_3\Big]\cr 
&- \sum_{k=1}^4\chi_{\delta,hk}\Big[n d_3 +\bar n {\bar d}_3+ 
d_1 d_2 +{\bar d}_1 {\bar d}_2\Big]
+ \sum_{k=1}^4\chi_{\alpha,hk}\Big[n {\bar d}_3 +\bar n {d}_3+ 
d_1 {\bar d}_2 +{\bar d}_1 {d}_2\Big]\cr
&- \sum_{k=1}^4\chi_{\delta,fk}\Big[n d_2 +\bar n {\bar d}_2+ 
d_1 d_3 +{\bar d}_1 {\bar d}_3\Big]
+ \sum_{k=1}^4\chi_{\alpha,fk}\Big[n {\bar d}_2 +\bar n {d}_2+ 
d_1 {\bar d}_3 +{\bar d}_1 {d}_3\Big]}
$$
and together with the direct channel M\"obius amplitude
$$
\cM_{3}^{(0)}=\h\Big(\chi_{\gamma,o1}-\chi_{\gamma,o2}-\chi_{\gamma,o3}
-\chi_{\gamma,o4}\Big)\Big[n+\bar n+\sum_{r=1}^3(d_r+\bar{d}_r)\Big]
$$
we get a vector and six scalars in the adjoint of the gauge group:
$${
{\rm U}(16)_9\otimes {\rm U}(16)_{5_1}\otimes {\rm U}(16)_{5_2}\otimes
{\rm U}(16)_{5_3}\ ,}
$$
and additional massless matter:
$$
\eqalign{
(\la_\alpha+\la_{\dot\alpha})\ &: \ (120\oplus {\overline {120}},1,1,1)\oplus
(1,120\oplus {\overline  {120}},1,1)\oplus 
(1,1,120\oplus {\overline {120}},1)\cr
& \oplus(1,1,1,120\oplus {\overline {120}})\cr
3(\la_\alpha+\la_{\dot\alpha})\ &: \ (136\oplus {\overline {136}},1,1,1)
\oplus (1,136\oplus {\overline
  {136}},1,1)\oplus (1,1,136\oplus {\overline {136}},1)\cr
&\oplus (1,1,1,136\oplus {\overline {136}})\ }
$$
in the untwisted sector, and 
$$
\eqalign{
(\la_\alpha+\la_{\dot\alpha})\ &: \ (16,16,1,1)\oplus
(1,1,16,16)\oplus 
(16,1,16,1)\oplus (1,16,1,16)\oplus (16,1,1,16)\cr
&\oplus (1,16,16,1)\oplus {\rm c.c.}\cr
2 \phi\ &:\ (16,{\overline {16}},1,1)\oplus (1,1,16,{\overline {16}})\oplus
(16,1,{\overline {16}},1)\oplus (1,16,1,{\overline {16}})\oplus (16,1,1,{\overline {16}})\cr
&\oplus (1,16,{\overline {16}},1)\oplus {\rm c.c.}\ }
$$
in the twisted sector.

In four dimensions one has to worry about gauge anomalies. However,
the common feature of all three models is that their spectrum is non-chiral
and therefore automatically free of gauge anomalies.

Since type 0B is non-supersymmetric from the very beginning,
it would be interesting to generalize this construction to  
$Z_N\times Z_M$ orbifolds that do not preserve any supersymmetry.

\bigbreak\bigskip\bigskip\centerline{{\bf Acknowledgements}}\nobreak
I am grateful to C. Angelantonj for very useful discussions.
This work was supported in part by EEC under the TMR contract 
ERBFMRX-CT96-0090.


\appendix{A}{}

\subsec{Combinations of ${\rm SO}(2)$ characters}

In the untwisted sector the expressions 
$\alpha_{ij}$,  $\beta_{ij}$, $\gamma_{ij}$ and $\delta_{ij}$  
with $i,j=o,g,h,f$ are: 
$$
\eqalign{
\alpha_{oo}&=O_2 O_2 O_2 O_2 +V_2 V_2 V_2 V_2\ ,\quad\quad
\alpha_{og}=O_2 O_2 V_2 V_2 +V_2 V_2 O_2 O_2\ ,\cr
\alpha_{oh}&=O_2 V_2 V_2 O_2 +V_2 O_2 O_2 V_2\ ,\quad\quad
\alpha_{of}=O_2 V_2 O_2 V_2 +V_2 O_2 V_2 O_2\ ,}
$$
$$
\eqalign{
\beta_{oo}&=O_2 V_2 V_2 V_2 +V_2 O_2 O_2 O_2\ ,\quad\quad
\beta_{og}=O_2 V_2 O_2 O_2 +V_2 O_2 V_2 V_2\ ,\cr
\beta_{oh}&=O_2 O_2 O_2 V_2 +V_2 V_2 V_2 O_2\ ,\quad\quad
\beta_{of}=O_2 O_2 V_2 O_2 +V_2 V_2 O_2 V_2\ ,}
$$
$$
\eqalign{
\gamma_{oo}&=S_2 S_2 S_2 S_2 +C_2 C_2 C_2 C_2\ ,\quad\quad
\gamma_{og}=S_2 S_2 C_2 C_2 +C_2 C_2 S_2 S_2\ ,\cr
\gamma_{oh}&=S_2 C_2 C_2 S_2 +C_2 S_2 S_2 C_2\ ,\quad\quad
\gamma_{of}=S_2 C_2 S_2 C_2 +C_2 S_2 C_2 S_2\ ,}
$$
$$
\eqalign{
\delta_{oo}&=S_2 C_2 C_2 C_2 +C_2 S_2 S_2 S_2\ ,\quad\quad
\delta_{og}=S_2 C_2 S_2 S_2 +C_2 S_2 C_2 C_2\ ,\cr
\delta_{oh}&=S_2 S_2 S_2 C_2 +C_2 C_2 C_2 S_2\ ,\quad\quad
\delta_{of}=S_2 S_2 C_2 S_2 +C_2 C_2 S_2 C_2\ .}
$$
Here the first factor refers to the two transverse 
space-time directions and the remaining three to 
internal coordinates. 
For the $g$-twisted sector, we get
$${\eqalign{
\alpha_{go}&=O_2 O_2 C_2 C_2 +V_2 V_2 S_2 S_2\ ,\quad\quad
\alpha_{gg}=O_2 V_2 S_2 C_2 +V_2 O_2 C_2 S_2\ ,\cr
\alpha_{gh}&=O_2 V_2 S_2 C_2 +V_2 O_2 C_2 S_2\ ,\quad\quad
\alpha_{gf}=O_2 V_2 C_2 S_2 +V_2 O_2 S_2 C_2\ ,}}
$$
$$\eqalign{
\beta_{go}&=O_2 V_2 S_2 S_2 +V_2 O_2 C_2 C_2\ ,\quad\quad
\beta_{gg}=O_2 O_2 C_2 S_2 +V_2 V_2 S_2 C_2\ ,\cr
\beta_{gh}&=O_2 O_2 C_2 S_2 +V_2 V_2 S_2 C_2\ ,\quad\quad
\beta_{gf}=O_2 O_2 S_2 C_2 +V_2 V_2 C_2 S_2\ ,}
$$
$$
\eqalign{
\gamma_{go}&=S_2 S_2 O_2 O_2 +C_2 C_2 V_2 V_2\ ,\quad\quad
\gamma_{gg}=S_2 S_2 V_2 V_2 +C_2 C_2 O_2 O_2\ ,\cr
\gamma_{gh}&=S_2 C_2 V_2 O_2 +C_2 S_2 O_2 V_2\ ,\quad\quad
\gamma_{gf}=S_2 C_2 O_2 V_2 +C_2 S_2 O_2 V_2\ ,}
$$
$$
\eqalign{
\delta_{go}&=S_2 C_2 V_2 V_2 +C_2 S_2 O_2 O_2\ ,\quad\quad
\delta_{gg}=S_2 C_2 O_2 O_2 +C_2 S_2 V_2 V_2\ ,\cr
\delta_{gh}&=S_2 S_2 O_2 V_2 +C_2 C_2 V_2 O_2\ ,\quad\quad
\delta_{gf}=S_2 S_2 V_2 O_2 +C_2 C_2 O_2 V_2\ ,}
$$
and for the $h$-twisted sector:
$$
\eqalign{
\alpha_{ho}&=O_2 S_2 S_2 O_2 +V_2 C_2 C_2 V_2\ ,\quad\quad
\alpha_{hg}=O_2 S_2 C_2 V_2 +V_2 C_2 S_2 O_2\ ,\cr
\alpha_{hh}&=O_2 C_2 C_2 O_2 +V_2 S_2 S_2 V_2\ ,\quad\quad
\alpha_{hf}=O_2 C_2 S_2 V_2 +V_2 S_2 C_2 O_2\ ,}
$$
$$
\eqalign{
\beta_{ho}&=O_2 C_2 C_2 V_2 +V_2 S_2 S_2 O_2\ ,\quad\quad
\beta_{hg}=O_2 C_2 S_2 O_2 +V_2 S_2 C_2 V_2\ ,\cr
\beta_{hh}&=O_2 S_2 S_2 V_2 +V_2 C_2 C_2 O_2\ ,\quad\quad
\beta_{hf}=O_2 S_2 C_2 O_2 +V_2 C_2 S_2 V_2\ ,}
$$
$$
\eqalign{
\gamma_{ho}&=C_2 O_2 O_2 C_2 +S_2 V_2 V_2 S_2\ ,\quad\quad
\gamma_{hg}=C_2 O_2 V_2 S_2 +S_2 V_2 O_2 C_2\ ,\cr
\gamma_{hh}&=S_2 O_2 O_2 S_2 +C_2 V_2 V_2 C_2\ ,\quad\quad
\gamma_{hf}=S_2 O_2 V_2 C_2 +C_2 V_2 O_2 S_2\ ,}
$$
$$
\eqalign{
\delta_{ho}&=S_2 O_2 O_2 C_2 +C_2 V_2 V_2 S_2\ ,\quad\quad
\delta_{hg}=S_2 O_2 V_2 S_2 +C_2 V_2 O_2 V_2\ ,\cr
\delta_{hh}&=C_2 O_2 O_2 S_2 +S_2 V_2 V_2 C_2\ ,\quad\quad
\delta_{hf}=C_2 O_2 V_2 C_2 +S_2 V_2 O_2 S_2\ ,}
$$
Finally for the $f$-twisted sector, we get
$$
\eqalign{
\alpha_{fo}&=O_2 C_2 V_2 S_2 +V_2 S_2 O_2 C_2\ ,\quad\quad
\alpha_{fg}=O_2 C_2 O_2 C_2 +V_2 S_2 V_2 S_2\ ,\cr
\alpha_{fh}&=O_2 S_2 O_2 S_2 +V_2 C_2 V_2 C_2\ ,\quad\quad
\alpha_{ff}=O_2 S_2 V_2 C_2 +V_2 C_2 O_2 S_2\ ,}
$$
$$
\eqalign{
\beta_{fo}&=O_2 S_2 O_2 C_2 +V_2 C_2 V_2 S_2\ ,\quad\quad
\beta_{fg}=O_2 S_2 V_2 S_2 +V_2 C_2 O_2 C_2\ ,\cr
\beta_{fh}&=O_2 C_2 V_2 C_2 +V_2 S_2 O_2 S_2\ ,\quad\quad
\beta_{ff}=O_2 C_2 O_2 S_2 +V_2 S_2 V_2 C_2\ ,}
$$
$$
\eqalign{
\gamma_{fo}&=S_2 O_2 C_2 V_2 +C_2 V_2 S_2 O_2\ ,\quad\quad
\gamma_{fg}=S_2 O_2 S_2 O_2 +C_2 V_2 C_2 V_2\ ,\cr
\gamma_{fh}&=S_2 V_2 S_2 V_2 +C_2 O_2 C_2 O_2\ ,\quad\quad
\gamma_{ff}=S_2 V_2 C_2 O_2 +C_2 O_2 S_2 V_2\ ,}
$$
$$
\eqalign{
\delta_{fo}&=S_2 V_2 S_2 O_2 +C_2 O_2 C_2 V_2\ ,\quad\quad
\delta_{fg}=S_2 V_2 C_2 V_2 +C_2 O_2 S_2 O_2\ ,\cr
\delta_{fh}&=S_2 O_2 C_2 O_2 +C_2 V_2 S_2 V_2\ ,\quad\quad
\delta_{ff}=S_2 O_2 S_2 V_2 +C_2 V_2 C_2 O_2\ .}
$$

\subsec{Characters}
The structure of the $64$ characters $\chi_{\alpha,ik},
\chi_{\beta,ik}, \chi_{\gamma,ik}$ and $\chi_{\delta,ik}$ is:
$$
\eqalign{
\chi_{\rho,{i1}}&=\rho_{io}A_i+\rho_{ig}B_i
+\rho_{ih} C_i+\rho_{if} D_i\ ,\quad
\chi_{\rho,{i2}}=\rho_{io}B_i+\rho_{ig}A_i
+\rho_{ih} D_i+\rho_{if} C_i\ ,\cr
\chi_{\rho,{i3}}&=\rho_{io}C_i+\rho_{ig}D_i
+\rho_{ih} A_i+\rho_{if} B_i\ ,\quad
\chi_{\rho,{i4}}=\rho_{io}D_i+\rho_{ig}C_i
+\rho_{ih} B_i+\rho_{if} A_i\ ,}
$$
where $\rho_{ij}=\alpha_{ij},\beta_{ij},\gamma_{ij}, \delta_{ij}$ 
with $i,j=o,g,h,f$ for the untwisted, the $g$-twisted, the $h$-twisted
and the $f$-twisted sector, respectively.
We defined the internal part for the untwisted sector
$$
\eqalign{
A_o&=\phi_o\phi_o\phi_o+\phi_v\phi_v\phi_v\quad\ ,\quad
B_o=\phi_o\phi_v\phi_v+\phi_v\phi_o\phi_o\ ,\cr
C_o&=\phi_o\phi_o\phi_v+\phi_v\phi_v\phi_o\quad\  ,\quad
D_o=\phi_o\phi_v\phi_o+\phi_v\phi_o\phi_v\ ,}
$$
for the $g$-twisted sector
$$
\eqalign{
A_g&=\phi_o\phi_s\phi_s+\phi_v\phi_c\phi_c\quad\ ,\quad
B_g=\phi_o\phi_c\phi_c+\phi_v\phi_s\phi_s\ ,\cr
C_g&=\phi_o\phi_s\phi_c+\phi_v\phi_c\phi_s\quad\ ,\quad
D_g=\phi_o\phi_c\phi_s+\phi_v\phi_s\phi_c\ ,}
$$
for the $h$-twisted sector
$$
\eqalign{
A_h&=\phi_s\phi_s\phi_o+\phi_c\phi_c\phi_v\quad\ ,\quad
B_h=\phi_s\phi_c\phi_v+\phi_c\phi_s\phi_o\ ,\cr
C_h&=\phi_s\phi_s\phi_v+\phi_c\phi_c\phi_o\quad\ ,\quad
D_h=\phi_s\phi_c\phi_o+\phi_c\phi_s\phi_v\ ,}
$$
and for the $f$-twisted sector
$$
\eqalign{
A_f&=\phi_s\phi_o\phi_s+\phi_c\phi_v\phi_c\quad\ ,\quad
B_f=\phi_s\phi_v\phi_c+\phi_c\phi_o\phi_s\ ,\cr
C_f&=\phi_s\phi_o\phi_c+\phi_c\phi_v\phi_s\quad\ ,\quad
D_f=\phi_s\phi_v\phi_s+\phi_c\phi_o\phi_c\ ,}
$$
and
$$
\phi_o+\phi_v={1\o\eta^2}\ ,\quad
\phi_o-\phi_v={\th_3\th_4\o\eta^2}\ ,\quad
\phi_s+\phi_c={\th_2\th_3\o 2 \eta^2}\ ,\quad
\phi_s-\phi_c={\th_2\th_4\o 2\eta^2}\ .
$$

\listrefs

\end